\documentclass[11pt]{article}
\usepackage{epsf}
\usepackage{amsmath}
\usepackage{graphics}
\usepackage{cite}

\renewcommand{\thefootnote}{\#\arabic{footnote}}
\begin{document}

\newcommand{\gtrsim}{ \mathop{}_{\textstyle \sim}^{\textstyle >} }
\newcommand{\lesssim}{ \mathop{}_{\textstyle \sim}^{\textstyle <} }

\newcommand{\rem}[1]{{\bf #1}}

\renewcommand{\thefootnote}{\fnsymbol{footnote}}
\setcounter{footnote}{0}
\begin{titlepage}

\def\thefootnote{\fnsymbol{footnote}}

\vskip .5in
\bigskip
\bigskip

\begin{center}
{\Large \bf Remarks on Dark Matter Constituents\\}
\vskip .1in
{\Large\bf with Many Solar Masses}

\vskip .45in

{\bf Claudio Corian\`o\footnote{email: claudio.coriano@le.infn.it} and 
P.H. Frampton\footnote{email: paul.h.frampton@gmail.com}}

\vspace{0.5cm}
{\em Dipartimento di Matematica e Fisica "Ennio De Giorgi"\\
 Universit\`a del Salento and INFN Lecce, \\
 Via Arnesano, 73100 Lecce, Italy}

\end{center}

\vskip .4in
\begin{abstract}
\noindent
Dark matter constituents of many solar masses will accrete normal matter
which emits X-rays that can be downgraded to microwaves which may
distort the precisely-measured black-body spectrum of the
Cosmic Microwave Background. However, it is
known from elsewhere that spherical models of accretion
vastly overestimate the amount accreted and consequently
the emitted X-rays.  
Therefore, exclusion plots based on
spherical accretion for the allowed fraction of the dark matter
versus the MACHO mass give upper limits on intermediate-mass
MACHOs which are too severe, sometimes by orders of magnitude.
\end{abstract}

\end{titlepage}

\renewcommand{\thepage}{\arabic{page}}
\setcounter{page}{1}
\renewcommand{\thefootnote}{\#\arabic{footnote}}

\newpage

\noindent
\section{Introduction}
The masses which have been discussed in the literature
for the constitutents of dark matter
cover a huge range of at least a hundred orders of magnitude from ultra-light axions \cite{CF}
($10^{-22}\, eV$) to very massive black holes ($10^{78} eV
\sim 10^{12}M_{\odot}$). In the present article we shall discuss only the
upper end of this range, above $20 M_{\odot}$. In particular, we are
interested in PIMBHs (Primordial Intermediate-Mass Black Holes) in
the mass range from $20 M_{\odot}$ to $2000 M_{\odot}$
because these are MACHOS which can likely be detected by 
microlensing experiments \cite{ChaplineExpt} in the foreseeable future, 
perhaps even 
before first light in 2022 of the LSST (Large Synoptic Survey Telescope).  \\
Such very massive MACHOs are best sought by microlensing experiments
as was demonstrated by the MACHO Collaboration\cite{Alcock} which used
the Mount Stromlo Observatory in Australia during the 1990s. They found
many light curves with durations up to 230 days corresponding to a
MACHO mass of up to approximately $20 M_{\odot}$. There was no known
motivation at the time of \cite{Alcock} to seek light curves of longer
duration. Now there is motivation\cite{PF,Chapline,ChaplineFrampton,PHFlong}
arising during the last few years from a reappraisal of the
dark matter candidates. The specific suggestion of \cite{PF} is that the
Milky Way dark halo contains billions of PIMBHs mostly in the
microlensing-sensitive range between $20 M_{\odot}$ and $2000 M_{\odot}$.\\
At present a microlensing experiment is underway which started in February
2018 and which should be able to discover the PIMBHs by the end of 2019.
If light curves can be established with durations of about two years,
corresponding to a MACHO mass about $100M_{\odot}$, it would provide
strong support for such a dark matter theory, and one would then continue to
seek higher mass MACHOs to obtain a more complete knowledge of the
PIMBH mass function.\\
The PIMBHs idea for dark matter constituents does not require any new physics
beyond the Standard Model and General Relativity. The microscopic candidates
for dark matter such as WIMPs and axions do require new physics. The WIMP
started out in 1977 as a sterile neutrino, in the hands of Hut\cite{Hut} and of
Lee and Weinberg\cite{LeeWeinberg}. The WIMP idea became very popular when
it was identified with a particle naturally appearing in electroweak supersymmetry
\cite{Goldberg}. The WIMP became canonised \cite{KolbTurner} as a serious
candidate for dark matter in the cosmology community. At present, two large
experiments \cite{XENON3,LZ} are under construction to search for them.\\
The history and popularity of the axion is similar to the WIMP. It emerged in
connection with solution of the strong CP problem, then needed to be modified
to avoid a contradiction with experiment. This led to the idea of an axion
in the mass range between $1\, \mu$eV and $1\, m$eV. \\
As these microscopic
candidates cease to be found, astrophysical candidates and the microlensing
proposal have become more central.\\
It is not as simple as that, however, because of stringent claims for
the upper limits on the numbers of PIMBHs coming from accretion
and the X-ray emission which can distort the CMB spectrum. The first
such claim was in \cite{ROM} which ruled out PIMBHs by orders
of magnitude but it was withdrawn \cite{JPO} when
it became clear that spherical accretion, as in the Bondi model \cite{Bondi},
was unreliable. Much more recently, in \cite{AK}, a similar exclusion
was calculated but again used the Bondi model of spherical
accretion. Such unreliable upper limits on PIMBHs have been widely
quoted in the dark matter community\cite{KITP}.
\section{ Spherical accretion}
The simplest model of accretion on to a black hole assumes spherical
symmetry and radial inflow of the accreting material. It was invented
in 1952 by Bondi\cite{Bondi} and was based on the simple differential equation
which was nothing more than the intellegent intuitive guess

\begin{equation}
\left( \frac{dM}{dt} \right) = \pi R^2 \rho v
\label{Bondi}
\end{equation}
where $M$ is the black hole mass, R is the Bondi radius
\begin{equation}
R = \left( \frac{2 M G} {c_S^2} \right),
\label{BondiRadius}
\end{equation}
$\rho$ is the ambient density and $v$ is the black hole velocity
or $v=c_S$ if $v < c_S$ where $c_S$ is the speed of sound.The Bondi radius is outside of the Schwarzschild radius and characterizes the region where the accretion takes place.\\
Substituting Eq.(\ref{BondiRadius}) into Eq.(\ref{Bondi}) and
assuming, as is usual, $v \sim c_S$ gives

\begin{equation}
\left( \frac{dM}{dt} \right) = 4 \pi \rho \left( \frac{G^2 M^2}{c_S^3} \right).
\label{Bondi2}
\end{equation}
Although this model seems reasonable and has been a staple problem
for students in astrophysics classes, in recent years it has been possible
in at least two cases to make the acid test: does Eq.(\ref{Bondi2}) give
results which agree with experiment? \\
To be fair, as originally presented \cite{Bondi}
it was never intended to be precisely accurate but it was meant to provide the correct 
order of magnitude. The purpose of this comment is to show that even this 
modest aim is not always the case,
and that Eq.(\ref{Bondi2}) can overestimate the amount of accretion
not by only a factor of two but even by a few orders of magnitude.\\
The questionable assumption is of course spherical symmetry and radial inflow.
Physical intuition suggests that the accretion process may well be far more chaotic
than implied by such a strong assumption.
This question becomes paramount in discussing the viability of dark
matter models \cite{PF} as discussed in the Introduction where the dark matter
constituents are PIMBHs and where spherical accretion has been widely assumed \cite{ROM, AK}
in making exclusion plots for the allowed fraction of the dark matter 
versus the MACHO mass. Let us therefore study two explicit examples,
the SMBHs (SuperMassive Black Holes) near the centres of Galaxy M87
and of the Milky Way.\\
\section{Two Examples where Spherical Accretion Studied}
In order to check whether the spherical accretion (Bondi) model is a reliable
approximation, we shall discuss two well-studied examples involving
supermassive black holes (SMBHs).

\subsection{SMBH in M87 Galaxy}
This SMBH is one of the nearest, except for Sag A* and M31 (Andromeda), to Earth at a distance of only 16.6 Mpc and has therefore
been one of the most intensely studied. Its mass is exceptionally large, 
about $6.6 \times 10^9 M_{\odot}$,
and hence it has a correspondingly huge Schwarzschild radius of 131AU, over three times the radius of 
Pluto's orbit.\\
The expected X-ray flux from M87 has been computed assuming spherical accretion
and has been measured observationally. The results are summarised in
\cite{M87,M87-2}. Both of these works conclude that the rate of accretion
at a distance of several times the Schwarzschild radius is orders of magnitude
{\it less} than the prediction by the Bondi model.

\subsection{ SMBH in Milky Way} 
This SMBH, named Sagittarius A* (or SagA* for short),
is of course by far the nearest to Earth at a distance of just 7.8 kpc. Its mass is exceptionally small fos a SMBH,
about $4 \times 10^6 M_{\odot}$ and its Schwarzschild radius is 12 million
kilometres, less than twice the radius of the Sun. This is quite different from the SMBH
in M87 and therefore provides a useful comparison.\\
The X-rays from SagA* have been predicted by spherical accretion
and also measured by direct observation. The X-rays
from Sag A* have strong variability and because this SMBH can be quiescent or flaring.
Nevertheless, Sag A* is yet another black hole accretion source
which is remarkably underluminous compared to the predictions
by the Bondi model. This has been suspected for twenty years
\cite{Quateart} and an improved treatment \cite{Balbus}
of the accretion involves vigorous circulation and outflow
in sharp contrast to the Bondi model. \\
A more recent survey of Sag A* is in \cite{Eckert}. The bottom
line is that we can observe X-rays from Sag A* at all only because
of its close proximity (8kpc) to Earth. Like the SMBH in M87
the Bondi model overestimates the X-rays by orders
of magnitude. 
Another careful discussion of Sag $A^*$ was made in \cite{Wang}
where it was found that the maximum luminosity is
only $\sim 10^{-9}$ of the Eddington limit. The peak lumosity during
flares is $\sim 5 \times 10^{35} ergs / s$. Sag A* has highly variable
emission ranging from flares with this maximum luminosity
and quiescent periods with very little X-ray emission. Nevertheless,
the Bondi prediction overestimates the X-ray luminosity by
between 4 and 8 orders of magnitude.
Although M87 and Sag A* are more massive than
the putative PIMBHs, it is reasonable to assume that using
the Bondi model of accretion for PIMBHs is comparably as
unreliable and will overestimate the mass accreted.\\
Let us reasonably assume that the Bondi model of accretion
is definitely {\it not} a good approximation for the PIMBH dark matter,
because of the reasons discussed in \cite{Balbus}.
This casts new light on the upper bounds
frequently quoted \cite{KITP} in the dark matter
community which exclude the idea that MACHOs in the
mass range $20 - 2000 M_{\odot}$ can be $100\%$ of
the halo dark matter, either for a monochromatc mass
or a smooth mass function.\\
For example the influential paper in 2008 by Ricotti, Ostriker
and Mack\cite{ROM} excluded PIMBHs above $10M_{\odot}$ by four orders of
magnitude using Bondi accretion. Eight years
later in 2016 the senior author of ROM conceded privately \cite{JPO} that
the ROM limits had been "far too severe". More recently, however.
in 2017, Ali-Ha\"imoud and Kamionkowski\cite{AK} presumably
unaware of the concession by ROM
published limits which similarly
rule out PIMBHs as all the halo dark matter. Yet the
accretion model employed in \cite{AK} was spherical and so
their conclusions are rendered highly questionable too.\\

\section{Reevaluation of CMB Distortion}

\noindent
One of the most striking facts about observational cosmology
is the astonishing agreement between the CMB (Cosmic Microwave
Background) spectrum and a Planckian black-body spectrum:
indeed the deviation is nearly too miniscule ($\lesssim 10^{-5}$)
to be measured. Naturally therefore the X-rays emitted
by PIMBH dark matter are of great concern in this respect.
However, we must be very careful in estimating the X-ray luminosity.
The main motivation of the present article is to point out that the
upper limits on PIMBHs in the literature \cite{ROM,AK} are too severe because
the calculations have used spherical accretion. This model was invented\cite{Bondi}
with simple, seemingly plausible, assumptions but not strictly derived
and, although Eq.(\ref{Bondi}) is an appealing
approximation, it simply does not describe the physics with sufficient
accuracy and can therefore be very misleading. As we have seen through the explicit
physical examples of M87 and SagA*, the Bondi model can overestimate
accretion by orders of magnitude relative to observations.\\
The masses of the black holes involved in M87 and SagA* are repectively
billions and millions of solar masses. The black holes being sought by
the microlensing experiment \cite{ChaplineExpt} are in the range
$20-2000 M_{\odot}$ but there is, to our knowledge, no reason to believe that their accretion
mechanism will be very different from that of the SMBHs.\\
Note that we discuss the range of PBHs all the way up to $10^{12} M_{\odot}$ which includes
the supermassive black holes at galactic centres \cite{PHFPBHlimit,AdSCFT,kong}.
Although the PIMBHs in the dark galactic halo must be below $10^6 M_{\odot}$,
because of disk stability arguments, we believe the primordial origin may be common 
to the halo dark matter, to the more massive black holes which
may exist as dark matter in clusters and to the galactic-core SMBHs themselves. \\
There are at least two large experiments, XENONnT\cite{XENON3}
and LZ\cite{LZ} starting to make an even more thorough search for
WIMPs, to follow up several unsuccessful earlier searches with less
sensitivity. While the WIMP retains some motivation, we should ensure
that the heaviest dark matter candidates have not been excluded
prematurely.

\section{Discussion}

Of course, a natural next step would be to do the accretion calculation
better and hopefully correctly. Since such analysis are quite involved and rely heavily on numerical simulations, here we just intend to draw the attention to two different significant effects which can contribute
to the large discrepancy with the Bondi model. \\
The first effect is the 
fact that the PIMBH spends the majority of its time in the
dark halo where there are few baryons to accrete and only
$\sim 1\%$ of the time inside the galactic disk where essentially
all of the baryons reside. Taking the typical thickness of the
disk as $t \sim 0.3\, kpc$ and the radius of a typical PIMBH orbit
as $R \sim 30\, kpc$ and given two transits through the disk per orbit
there is a crude estimate $2 t / 2 \pi R = t/\pi R \sim 3 \times 10^{-3}$
of the time spent in a baryonic environment. Certainly assuming 
a time-independent density of the environment in Eq. (\ref{Bondi}) is disfavored.\\
A second effect absent from the Bondi model
(which might work better for the case of a static Schwarzschild black hole)
is due to the inevitable Kerr-like large angular momentum of the
PIMBH which renders accretion on to such a Kerr black hole
more difficult than for a stationary
Schwarzschild black hole. As discussed in {\it e.g.} \cite{JofBHs}
the pin angular momentum of a $100M_{\odot}$ can be
typically be $\sim 500$ times that of the Sun. What is needed to accommodate this
second effect could be an angular-momentum-barrier correction in Eq.(\ref{Bondi})
and we intend to study this second effect in a forthcoming work.

\section*{Acknowledgement}

\noindent
We thank S. Balbus for a useful discussion. This work is partially supported by INFN INiziativa Specifica QFT-HEP.
P.H.F. thanks the Physics Department at the University of Salento for hospitality.


\vspace{0.5in}



\begin{thebibliography}{100}
\bibitem{CF}C. Corian\`o and P.H. Frampton, {\it Dark Matter as Ultralight Axion-Like particle in $E_6\times U(1)_X$ GUT with QCD Axion}, 
Phys. Lett. B782, 380 (2018).
\bibitem{ChaplineExpt}
W. Dawson, {\it et al},
{\it Microlensing Experiment Using Blanco 4m Telescope, Cerro Tololo, Chile.}
Started February 2018.
\bibitem{Alcock}
C. Alcock, {\it et al.}, (The MACHO Collaboration),
{\it The MACHO Project: Microlensing Results fron 5.7 Years of LMC Observations}.
Astrophys. J. {\bf 542,} 281 (2000).  {\tt arXiv:astro-ph/0001272}.
\bibitem{PF}
P.H. Frampton,
{\it Searching for Dark Matter Constituents with Many Solar Masses.}
Mod. Phys. Lett. {\bf A31,} 1650093  (2016).  {\tt arXiv:1510.00400[hep-ph]}.
\bibitem{Chapline}
G.F. Chapline,
{\it Cosmological Effects of Primordial Black Holes}.
Nature {\bf 253,} 251 (1975).
\bibitem{ChaplineFrampton}
G.F. Chapline and P.H. Frampton,
{\it A New Direction for Dark Matter Research: Intermediate Mass Compact Halo Objects}.
JCAP {\bf 1611:}042 (2016).  {\tt arXiv:1608.04297[gr-qc]}
\bibitem{PHFlong}
P.H. Frampton,
{\it On the Origin and Nature of Dark Matter}.
{\tt arXiv:1804.03516[physics.gen-ph]}
\bibitem{Hut}
P. Hut,
{\it Limits on the Masses and Number of Weakly Interacting Particles}.
Phys. Lett. {\bf B69,} 85 (1977).
\bibitem{LeeWeinberg}
B.W. Lee and S. Weinberg,
{\it Cosmological Lower Bound on Heavy NeutrinoMasses}.
Phys. Rev. Lett. {\bf 39,} 165 (1977).
\bibitem{Goldberg}
H. Goldberg,
{\it Constraints on the Photino Mass from Cosmology}.
Phys. Rev. Lett. {\bf 50,} 1419 (1983).
\bibitem{KolbTurner}
E.W. Kolb abd M.S. Turner,
{\it The Early Universe}
Frontiers in Physics. (1990).
\bibitem{XENON3}
E. Aprile, {\it et al.} (XENON Collaboration),
{\it The XENON1T Dark Matter Experiment}.
Eur. Phys. J. {\bf C77,} 881 (2017).
{\tt arXiv:1708.07051[astro-ph.IM]}.
\bibitem{LZ}
The LZ Dark Matter Experiment.
{\tt lz.lbl.gov}.
\bibitem{ROM}
M. Ricotti, J.P. Ostriker and K.J. Mack, 
{\it Effect of Primordial Black Holes on the Cosmic Microwave Background 
and Cosmological Parameter Estimates.} 
Astrophys. J. {\bf 680,} 829 (2008). 
{\tt arXiv:0709.0524[astro-ph]}.
\bibitem{JPO}
J.P.Ostriker, 
Private communication. (December 28, 2016).
\bibitem{AK}
Y. Ali-Ha\"imoud and M. Kamionkowski,
{\it Cosmic Microwave Background Limits on Accreting Primordial Black Holes}.
Phys. Rev. {\bf D95,} 043534 (2017). {\tt arXiv:1612.05644[astro-ph.CO]}
\bibitem{KITP}
http://online.kitp.ucsb.edu/online/cdm-c18/
\bibitem{Bondi}
H. Bondi,
{\it On Spherically Symmetric Accretion}.
MNRAS {\bf 112,} 195 (1952).
\bibitem{M87}
C.Y. Kuo, K. Asada, R. Rao, M. Nakamura, J.C. Algaba, H.B. Liu, M. Inoue, P.M. Koch, P.T.P. Ho, S. Matsushita, H.-Y. Pu, K. Akiyama, H. Nishioka and N. Pradel, 
{\it Measuring Mass Accretion Rate onto the Supermassive Black Hole in M87 Using Faraday Rotation Measure with the Submillimeter Array}.
Astrophys. J. {\bf 783,} L33 (2014). 
{\tt arXiv:1402.5238[astro-ph.GA]}.
\bibitem{M87-2}
H.R. Russell, A.C. Fabian, B.R. McNamara and A.E. Broderick,
{\it Inside the Bondi Radius of M87}.
MRNAS {\bf 451,} 588 (2015). 
{\tt arXiv:1504.07633[astro-ph.GA]}.
\bibitem{Quateart}
E. Quateart, R. Narayan and M.J. Reid,
{\it What is the Accretion Rate in Sagittarius A*?}
Ap.J. Lett. {\bf 517,} L101 (1999).
{\tt arXiv:astro-ph/9903412}.
\bibitem{Balbus}
J.F. Hawley and S. Balbus,
{\it A Dynamical Structure of Nonradiative Black Hole Accretion Flows}.
Astrophys. J. {\bf 573,} 738 (2002).
{\tt arXiv:astro-ph/020309}.
\bibitem{Eckert}
A. Eckert, {\it at al.,}
{\it The Milky Way's Supermassive Black Hole}.
Found. Phys. {\bf 47,} 553 (2017).
{\tt arXiv:1703.09118[astro-ph.HE]}.
\bibitem{Wang}
Q.D. Wang, {\it et al.}\\
{\it Dissecting X-Ray Emitting Gas Around the Centre of Our Galaxy}.\\
Science {\bf 341,} 981 (2013).
{\tt arXiv:1307.5845[astro-ph.HE]}.
\bibitem{PHFPBHlimit}
P.H. Frampton,
{\it Maximum PBH Mass and Primordiality}.
{\tt arXiv:1802.05678[gr-qc]}
\bibitem{AdSCFT}
P.H. Frampton,
{\it AdS / CFT string duality and conformal gauge field theories}.
Phys.Rev. {\bf D60,} 041901 (1999). 
{\tt arXiv:hep-th/9812117}.
\bibitem{kong}
P.H. Frampton and O.C.W. Kong,
{\it Horizontal symmetry for quark and squark masses in supersymmetric SU(5)}.
Phys.Rev.Lett. {\bf 77,} 1699 (1996). 
{\tt arXiv:hep-ph/9603372}.
\bibitem{JofBHs}
P.H. Frampton,\\
{\it Angular Momentum of Dark Matter Black Holes}.\\
Phys. Lett. {\bf B767,} 303 (2017).
{\tt arXiv:1608.04297[hep-ph]}.
\end{thebibliography}
\end{document}